\documentclass[12pt,pra,aps,amssymb,amsfonts,amsmath,tightenlines,nofootinbib]{revtex4}

\usepackage[dvips]{graphicx}
\usepackage{dsfont, color,txfonts} 
\usepackage{tcolorbox}
\usepackage{wrapfig}

\newcommand{\re}{\mbox{$\rm e$}}
\newcommand{\rd}{\mbox{$\rm d$}}
\newcommand{\half}{\mbox{$\textstyle \frac{1}{2}$}}

\begin{document}

\title{Mathematical Politics}
\author{Dorje C. Brody}

\affiliation{
School of Mathematics and Physics, University of Surrey, 
Guildford GU2 7XH, UK, and  
\\ Department of Mathematics, Imperial College London, London SW7 2BZ, UK  
}

\date{\today}

\begin{abstract}
\noindent 
Politics today is largely about the art of messaging to influence the public, but the 
mathematical theory of messaging --- information and communication theory 
--- can turn this art into a precise analysis, both qualitative and quantitative, 
that enables us to gain retrospective understandings of past political events and 
to make forward-looking future predictions. 
\end{abstract}

\maketitle


\section*{Introduction}

Over the years I have observed that some of my colleagues in mathematical 
science community are oblivious to what is happening in politics. Perhaps 
this is a healthy attitude, given all that is taking place around the world today, 
but for others who are more in tune with politics, it might be a revelation that 
the power of mathematical analysis stretches into politics to make concrete 
statements. Mathematical politics, or `political mathematics', is not about 
politics taking place within the mathematical science community (about which 
those colleagues do have much to say!). Rather it concerns mathematical 
modelling that enables us to understand political events, such as elections 
or referendums, and in some cases allows us to make future 
predictions.\footnote{I am not concerned here with mathematical approaches 
to social choice in political economy (Arrow 1963, Taylor 1971), nor with any 
other areas of mathematical economics that may have implications in 
policy choice. The term `mathematical politics' has indeed been used once 
in the context of Arrow's impossibility theorem (Samuelson 1967). Today's 
politics, however, has diverged away from the traditional idea that 
mathematical economics might assist politicians in designing better economic 
policies. Rather, with the advances in communication technology, today's 
politics has reduced largely to the art of messaging, often lacking in substance 
for ideas to tackle real societal challenges, and this is where the science of 
information, communication, and control can play significant roles, both in 
positive and negative ways.} 

The word politics carries multiple meanings, but broadly speaking I 
have in mind here the activities of the government or lawmakers to try to 
influence the way a country is governed. A country can be replaced by an 
organisation, in which case we will be speaking of an organisational politics, 
but let me stay with the governance of a country in a broadly democratic 
system. Then to implement a policy, one must either convince the 
government to take your advice, or else convince the voters to allow you form 
a government. From this point of view, politics boils down to convincing others 
through messaging. If we want a scientific approach to this process, then we 
must understand how people digest messages and extract information 
contained therein. This is where we can apply 
powerful mathematical techniques of information and communication theory 
with the hope of making a difference, and that is the topic I would like to outline 
in what follows. I begin by discussing closely-related ideas in financial 
modelling, which provides a context in which it is easier to understand 
the role of information in society. Then I proceed to discuss the calculus for 
elections, and explore other ideas that are important in political messaging.

\section*{A question arising from financial modelling}

Norbert Wiener, arguably one of the greatest mathematicians to have been 
born in America, made the point in the late 1940s that society can only be 
understood through the lens of communication theory (Wiener 1950), 
because communication is the basis of society. Although Wiener made a 
broad range of predictions on the future forms of society, many of which, 
including one on the environmental challenge (box below), 
became reality, he did not implement any concrete mathematical model 
to study any particular social system.

\begin{tcolorbox}

\section*{From {\em The Human Use of Human Beings} (Wiener 1950)}
\vspace{-0.4cm} 
``This is partly the result ... of an increased mastery over nature which, on a limited 
planet like the earth, may prove in the long run to be an increased slavery to 
nature. For the more we get out of the world the less we leave, and in the long run 
we shall have to pay our debts at a time that may be very inconvenient for our own 
survival. ... We have modified our environment so radically that we must now 
modify ourselves in order to exist in this new environment.'' 

\end{tcolorbox}

Neither did anyone else implement Wiener's vision in the fifty years after his 
book of 1950 in the context of social science, although the mathematical 
concepts 
underpinning information and communication theory developed rapidly in 
areas such as coding theory, signal processing, and stochastic filtering, all of 
which have significant practical applications in various areas of engineering 
and technology. 

Only in the past two decades, a new and unexpected application emerged 
that embodies Wiener's philosophy in the context of financial modelling. In the 
financial industry, the pricing and risk management of financial contracts was a big 
issue twenty to thirty years ago. These financial contracts, known as contingent 
claims, 
entail random payouts that are dependent on the future prices of one or 
more underlying assets. An example of such an asset is a stock. The pricing issue is 
then not concerned with the price of the stock, which is determined by the market, 
but the pricing of a contract whose payout is a function of the future stock price. 

Following the tradition of mathematical modelling, one thus requires a model for 
the price of the stock, from which the prices of contingent claims on that stock 
can be worked out. This is natural inasmuch as we observe the daily ups and 
downs of the stock price, so we construct a model for what we see, in the form of 
a solution to a stochastic differential equation. But is it really the right approach, if 
we want to obtain logically motivated asset-price models? 

\section*{Understanding information flow is the key}

The issue here is that the daily movements of asset prices result as a 
consequence of the daily revelation of information. Yet, if we start out from a 
model for the asset price, it is not possible to connect that model to the 
information that generated  the price movements in the first place. In other words, 
the causal relation between the flow of information and the price dynamics is lost. 
It is then inevitable that the chosen model for the stock price will have little to do 
with the actual price movements. So there has to be a big leap of imagination to 
do better. But how? 

For a long time, economists have understood that the main driving factor of 
price movements 
in financial markets is information. However, unlike price movements of a stock 
that can be measured as a time series, information, such as news bulletins, does 
not come in the form of a numerical 
time series. The intuition about the impact of information on the price is clear 
enough --- we read a news article about global protests of Tesla cars and we 
expect the Tesla stock price to drop. But this news article is 
presented in the form of words and sentences. 
So how can we model such 
information mathematically? 

The answer is to rely on the notion of a `random variable' in probability. A random 
variable assigns a number to an outcome of chance, and is an indispensable tool 
for modelling the statistics of random phenomenon. An example is the outcome 
of coin toss, which is either a head or a tail. If we wish to apply statistical analysis 
to the outcomes of a fair coin, such as the mean of the outcome, we encounter an 
issue because we cannot average a head and a tail --- there is no 
`Cecrops' face of a coin. But if we assign the 
numerical value zero to a tail and one to a head, say, 
which is what a random variable 
does, then we can meaningfully say that the average outcome of a fair coin is a 
half. 

The idea is to extend this notion to general information that we digest. For this 
purpose, the first step is to identify the relevant quantity of interest contained in a 
message, and associate numerical values to label different realisations of this 
quantity. For example, 
in a financial investment, the question an investor would ask is: What is 
the return? Needless to say, no one is actively transmitting this information, 
unlike in a conventional communication channel. Nevertheless, we can 
interpret the quantity of interest as the signal component 
of a message in communication theory. Of course, information is rarely perfect in 
financial markets --- no one knows the future price of an asset. The imperfection 
of information, 
representing rumours, speculations, or misinterpretations, can 
then be modelled in the form of noise in a communication channel. Putting these 
ideas into a simple formula, the acquisition of noisy information on the future 
payout $X$ of an investment can be translated into the detection of the value of 
the noisy information 
\[
\xi = \sigma X + \epsilon \, , 
\] 
where $\epsilon$ represents noise, modelled by a random variable independent 
of $X$, and I have also inserted the signal-to-noise ratio $\sigma$. 

In the case of a financial investment, the modelling of the signal 
is straightforward because 
the quantity of interest, namely the return, is already expressed numerically. In 
general this is not the case, and this is where one has to be creative in assigning 
numerical values to different realisations of the quantity of interest as a random 
variable. For example, if one is concerned with making a choice among $N$ 
alternatives, then these alternatives can be labelled by $N$ numbers $x_1$, 
$x_2$, $\cdots$, $x_N$. Let $p_k$ be the current likelihood of alternative 
$x_k$ being selected. Then we have a model for the signal $X$. For a binary 
choice we thus have $X=x_1$ or $X=x_2$. Any speculative information (noise) 
that suggests that the correct choice is $X=x_2$ is represented by the random 
variable $\epsilon$ taking a large value --- the larger the value is, the stronger 
the suggestion is, and conversely smaller values of $\epsilon$ represents 
speculations towards the alternative $X=x_1$. If there are many such 
speculative or misguided 
opinions, then the law of large number suggests that $\epsilon$ may 
be modelled by a zero-mean Gaussian random variable. An information model 
is then constructed. 

Although the 
mathematics here is simple, the idea it conveys is far reaching, because it 
goes against our instincts for mathematical modelling: Rather than modelling the 
price dynamics that we do observe in numbers, we model news items, company 
reports, central bank announcements, 
and so on, that we might read on the Internet. So does it work? In financial 
modelling 
the proof of the pudding is in the `calibrating' -- 
the process of adjusting model parameters 
to fit the market prices of contingent claims. For a practical application in 
financial modelling, though, the 
arrival of information must be modelled using a time series, for example, 
\[
\xi_t = \sigma X t + \epsilon_t \, , 
\] 
where we let the noise process $\{\epsilon_t\}$ be modelled by a Brownian 
motion. When an information model of this kind is applied to generate 
models to price associated 
contingent claims, the fit to market data was better than the 
ones used by many of the mainstream investment banks, importantly 
with fewer model parameters (Mackie 2011). 
Of course, it is not possible to accurately model the complicated way 
in which information is spread and 
digested in the market, but by modelling the root of the 
price dynamics, one can generate rather realistic models for price 
processes. As 
George Box famously put it, `all models are wrong'; but these information models 
are certainly useful. We called this framework an information-based asset 
pricing (Brody \textit{et al}. 2008).

\section*{Innovation: the arrival of new information}

When we digest a `news' article on the Internet or in the papers, often much of 
the contents in the article is not new. News editors often modify an older article 
to include the latest developments so that the article is reasonably 
self-contained. This is indeed quite common --- when we acquire information, 
typically it contains both the known and the unknown. Importantly, though, we 
tend to respond to the unknown bits because they represent the arrival of 
genuinely new information. 

In communication theory there is a standard procedure to isolate the arrival of 
new information from an information-providing time series $\{\xi_t\}$. 
To understand this, 
consider the information gathered over the time interval $[t, t+\rd t]$, which is 
represented by the increment $\rd\xi_t$. If we write ${\mathbb E}_t[-]$ for the 
expectation conditional on the information available up to time $t$, then the 
predicted arrival of new information at time $t$ is given by 
${\mathbb E}_t[\rd\xi_t]$. It then follows that the discrepancy between the 
predicted and realised increments 
\[
\rd W_t = \rd\xi_t - {\mathbb E}_t[\rd\xi_t] 
\] 
represents the surprise element, that is, the arrival of new information. The 
process $\{W_t\}$ defined by this relation, which in itself is also a Brownian 
motion but is rather distinct from the noise, is known as the `innovation process'. 
According to Thomas Kailath, both the concept and 
terminology were introduced by Wiener (Kailath 1968). 

The significance of the innovation process in financial modelling and beyond is 
the following. The fundamental theorem of asset pricing asserts that the price at 
time $t$ of an asset entailing a future payout $X$ is given by the expectation 
${\mathbb E}_t[X]$, suitably discounted, in a certain probability measure known 
as the `risk-neutral probability', 
conditional on the information available up to time $t$. In the conventional 
approach to mathematical modelling, as indicated earlier, one specifies a model 
for the asset price in the form of a stochastic differential equation, driven by a 
Brownian motion that represents market uncertainties, and assumes that 
the solution to this equation generates market information, as well as the future 
return $X$. The hidden implication, not stated in textbooks, is that markets are 
assumed to be driven by uncertainties. 

Markets have views on the future return of an asset, and they are not 
based on which stochastic differential equation a modeller might have used. 
These views change in response to 
new information, leading to dynamics; whereas in the traditional approach, the 
assumption is that markets obtain information through observing the asset 
price dynamics. The causality structure is therefore flipped around. This is why 
in the information-based approach we model the return $X$ along with how 
information flows to inform markets about the possible values of this return. 
The price process is then generated through an expectation ${\mathbb E}_t[X]$, 
not by solving a differential equation. But having deduced a model for the asset 
price in this way, one can then ask which differential equation it satisfies. The 
result reveals that the price process is driven by the innovation process 
$\{W_t\}$, that is, the arrival of new information, and not by market uncertainties 
$\{\epsilon_t\}$; and this is the significant fact about the information-based 
approach that extends beyond mere financial modelling.

\section*{Structural modelling of electoral competition}

There are many systems whose dynamics are driven by information, 
not merely in financial markets. With the advances in artificial intelligence 
and increased concerns 
on the impacts of disinformation in society, it will be irresponsible for scientific 
communities to sit back and pretend that these issues do not concern them, 
especially when they are equipped with the powerful machinery of the calculus 
of information. Since the primary concern of the impact of disinformation lies 
in politics and democratic processes, we need a modelling framework for 
these processes. The information-based asset pricing theory tells us exactly how 
this can be done. 

As an example, consider the modelling of 
an electoral competition. In an election there are a 
handful of key factors that are of relevance to a large number of electorates. 
These are the policy positions that would be taken by the candidates if they 
were elected, 
on taxation, public spending, abortion rights, gun control, climate policy, 
the Middle East, and 
so on. They may also include the personality or charisma of the candidates. 
Voters are unaware of which position a given candidate might take on any 
given issue, if elected, but have partial information about this. We let a 
random variable $X_k^m$ model the position of candidate $m$ on policy $k$. 

Political messages will either be concerned with convincing electorates that their 
policy positions are advantageous to them, or saying whatever they think the 
electorates want to hear. Some messages will focus on convincing the electorates 
how the perceived 
policy positions of other candidates will be undesirable. In one way or another, 
there will be a lot of noise. 

For each election factor $X^m_k$, we thus have a time series $\{\xi_t^{m,k}\}$ of 
noisy information, the aggregate of which for all $m$ and all $k$ determines 
the information available to the electorates, which they can use to arrive at 
their best estimates for these factors. Voters will have varied preferences --- for 
instance, a position in support of stricter gun control may be perceived as good for 
one voter but bad for another. Voter preferences will be modelled as weighting 
factors that can be sampled from a distribution over voter preferences. Then for 
each voter we can assign scores for the candidates, with the rule that the 
candidate with the highest score on the election day will be selected by that 
voter. This is the idea of a structural model for an election (Brody \& Meier 2002), 
which can be implemented in different practical situations.

\section*{Calculus for elections: a reduced-form approach}

Although a structural model has advantages in micro-managing 
political messages during a major election campaign, qualitative 
features of such a model can be captured using a somewhat simplified 
reduced form model discussed now. Here we let 
a single random variable $X$ represent different 
candidates. Thus, if there are $n$ candidates then $X$ takes the values 
$x_1,x_2,\ldots,x_n$, labelling the candidates, 
with the \textit{a priori} probabilities $p_1,p_2,\ldots,p_n$. 
These probabilities have the interpretation of representing the current poll 
statistics. 

To understand the need of a dynamical model, consider an election involving 
two candidates, $A$ and $B$. Suppose that today's poll shows that candidate 
$A$ has 52\% support, and candidate $B$ has 48\%; and the election is 
to take place in one year. Ignoring errors in poll statistics, and assuming a 
first past the post election system, what do these numbers 
tell us? What do they say about the likelihoods of the candidates winning the 
election? 

What these numbers tell us is that if an election were to take place 
tomorrow, then the probability of candidate $A$ winning the election is about 
100\%. But they say very little about what happens in a future election. If the 
election were to take place in a year, then most people will think that the 
likelihood of candidate $A$ winning the election is close to 52\%. But what if 
the election is to take place in six months, or in three months? To 
interpolate today's statistics with future statistics, we need a dynamical model. 
In the present context, what generates the dynamics is the revelation of 
information, so let me take the simplest nontrivial model $\xi_t=\sigma X t + 
\epsilon_t$ for the flow of information. Then we can generate a model for the 
dynamics of the poll statistics by the conditional probability 
$\pi_{it}={\mathbb P}(X=x_i|\xi_{\leq t})$, where 
$\xi_{\leq t}$ denotes information gathered up to time $t$. Using the Bayes 
formula we find, in this example, 
\[ 
\pi_{it} = \frac{p_i \re^{\sigma x_i \xi_t - \frac{1}{2}\sigma^2 x_i^2 t}}
{\sum_i p_i \re^{\sigma x_i \xi_t - \frac{1}{2}\sigma^2 x_i^2 t}}  
\] 
for the \textit{a posteriori} support rates. This is how we can use a model for 
the flow of information to interpolate between today and the future. If the election 
is to take place at a future date $T$, then the vote share of candidate $j$ 
exceeding a threshold value $K$ on the election day, based on the knowledge 
today and on how information flows to the electorates from today till the election 
day, can be determined by working out the probability 
${\mathbb P}(\pi_{jT}>K)$. As an example, 
in the case of a two-candidate race (Brody 2019), 
writing $p_1=p$ and $p_2=1-p$, a calculation shows that 
\[ 
{\mathbb P}\left(\pi_{1T}>K\right) = p\, N(d^-) + (1-p)\, N(d^+) \, , 
\]
where 
\[
d^\pm =  \frac{ \log\left( \frac{(1-p)(1-K)}{pK}\right) \pm \frac{1}{2}\sigma^2 T}
{\sigma \sqrt{T}} 
\]
and $N(x)$ denotes the cumulative normal distribution function. It is a curious 
coincidence that the formula emerging from an information-based model for 
the projected probability ${\mathbb P}\left(\pi_{1T}>1/2\right)$, 
as of today, of a candidate winning a future election 
should coincide essentially with the famous Black-Scholes option pricing 
formula in asset pricing theory.

\begin{wrapfigure}{r}{0.50\textwidth}
\includegraphics[width=0.95\linewidth]{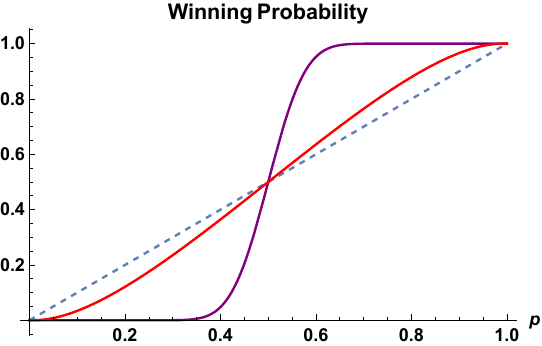} 
\caption{The probability of winning an 
election in one year, as a function of today's support rate $p$, for 
$\sigma=0.2$ (purple) and $\sigma=1.2$ (red). When $p>0.5$, 
the realised probability of winning a two-candidate election in the 
future is always higher than $p$, and conversely for $p<0.5$.}
\vspace{-0.5cm} 
\label{fig:wrapfig}
\end{wrapfigure}

There are lessons to be learned even from this simple model. If today's polls 
were indicators of the likelihoods of candidates winning a future election, a 
common misconception, then we would have ${\mathbb P}\left(\pi_{1T}>
1/2\right)=p$ for a two-candidate race. In reality the projected 
winning probability is 
always higher than today's support rate $p$ if $p>1/2$; and always lower 
than $p$ if $p<1/2$. Exactly how much higher (or lower) depends on the 
information flow rate $\sigma$. For instance, suppose that one candidate has 
52\% support rate today, and that $\sigma\ll1$. Then the realised probability 
of this candidate winning an election in a year is close to 1, because if 
that candidate has 52\% support today, and if voters learn very little over the 
coming year because $\sigma\approx0$, 
then those 52\% of voters will still vote for this candidate, so 
the probability of winning is close to 1. 
This line of analysis shows that if you are leading in a two-candidate race 
then you want very little information to come out; whereas if you 
are lagging behind then you want as much information as possible to come 
out. In other words, why rock the boat when you are leading? This common 
phrase one hears from political commentators during an election 
cycle can be justified on account of a simple analysis in mathematical politics. 

More generally, the rate $\sigma$ at which reliable information is revealed varies 
over time. In this case we lose the Markov property that the posterior probabilities 
are functions of the terminal value of the time series $\{\xi_t\}$. Nevertheless it is 
possible to work out the projected probabilities of the candidates winning a future 
election in closed form. By means of the method of functional derivative we can  
study optimal strategies for political messaging so as to maximise the 
winning probability (Brody 2019). 

When there are three or more candidates in a race, such an analysis becomes 
nontrivial but nevertheless is tractable (Brody \& Yuasa 2023). 
While in a two-candidate 
race the only control parameter is the signal-to-noise ratio $\sigma$, in an 
election with multiple candidates, their \textit{political spectrum gaps} $x_k-k_l$ also 
become control parameters. What this means is that we can implement a 
sensitivity analysis to study how repositioning of the candidates or their political 
parties in the political spectrum will affect the winning probabilities. For example, 
suppose that there are three parties, positioning themselves at centre left, 
centre right, and right, and that the right-leaning party is ranked second in 
today's poll. In this scenario, we find that if the right-leaning party moves 
further to the right, then their winning probability almost certainly drops, unless 
if $\sigma\ll1$, that is, unless if the election is overshadowed by confusion and 
uncertainty. We thus encounter here another strength of mathematical 
analysis that 
explains the empirically observed fact that rightwing political parties tend to 
disseminate a lot of disinformation (T\"ornberg \& Chueri 2025).

\section*{Tenacious Bayesians}

Let me now discuss an important dynamical characteristic of the Bayesian 
updating leading to the transformation $p_i\to\pi_{it}$ from the prior to the 
posterior. For this purpose it will 
be useful to take the derivative of $\pi_{it}$, making use of the Ito stochastic 
calculus. Specifically, $\pi_{it}$ is a function of two variables $t$ and $\xi_t$, 
so its (stochastic) derivative $\rd\pi_{it}$ can be worked out by expanding it as 
a Taylor series and retain terms of first order, bearing in mind that $(\rd\xi_t)^2 
= \rd t$. If we then substitute $\rd\xi_t=\rd W_t+{\mathbb E}_t[\rd\xi_t]$ into 
the result, we find that $\rd\pi_{it}\propto\rd W_t$. That is, increments of the 
posterior probability are proportional to those of the innovation process. Such 
a process is known as a `martingale', representing a conserved quantity, 
albeit on average. 

\begin{wrapfigure}{r}{0.50\textwidth}
\includegraphics[width=0.95\linewidth]{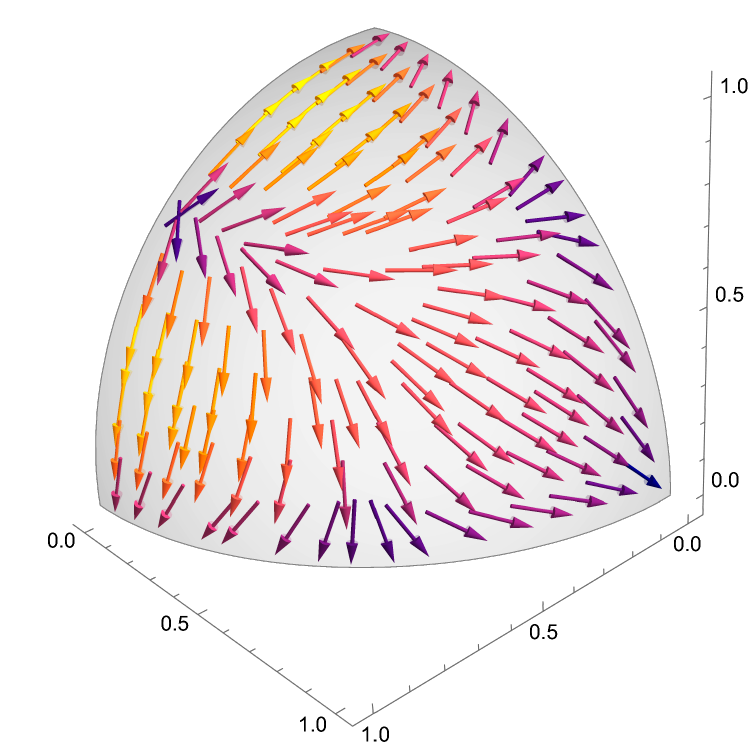}
\caption{Negative gradient flow of the variance. 
The three corners of the octant, where the flow attempts to take you to, 
correspond to states with zero uncertainty. }
\label{fig:2}
\end{wrapfigure}

The stochastic differential equation satisfied by $\{\pi_{it}\}$ is known as 
the `Kushner equation', which, in the present example is just an 
equation for a martingale, having no drift. To gain a deeper intuition of its 
dynamical characteristics, it will be helpful to bring in a geometric consideration. 
To proceed, consider the square root $\psi^i_{t}=\sqrt{\pi_{it}}$ of the posterior 
probability. Then squares of $\psi^i_{t}$ add up to unity, that is, $\{\psi^i_{t}\}$ 
determines a point on a unit sphere. 
The radius of this sphere, the total probability, however, is purely a matter of 
convention --- we can say that the likelihood of an event is 0.6, or 6 out of 10, or 
60\%; they all convey the same notion. Statistically relevant information is thus 
encoded in the direction of the vector $\psi^i_{t}$, not the length. 
The space of directions in the vector space on which $\{\psi^i_t\}$ is defined is 
known as a `real projective space', which is a compact Riemannian manifold 
${\mathfrak M}$. The dynamical equation for $\{\pi_{it}\}$ can then be lifted 
into a covariant differential equation (box below) on ${\mathfrak M}$, given by 
\[ 
\rd\psi^i_t = -\textstyle{\frac{1}{16}} \nabla^i V_t \, \rd t + 
\textstyle{\frac{1}{4}} \nabla^i X_t \, \rd W_t . 
\] 
Here, $X_t(\psi)$ 
is the conditional mean and 
$V_t(\psi)$ 
is the conditional variance of the 
random variable $X$; both are functions on ${\mathfrak M}$. The drift 
vector field of the state $\psi^i_t$, which determines the orientational 
direction, when there are three candidates, is plotted in figure~\ref{fig:2}. 

\begin{tcolorbox}

\section*{Covariant Ito derivative (Hughston 1996, Brody \& Hughston 2001)}
\vspace{-0.4cm} 
An \textit{Ito process} $\{x_t\}$ is defined by 
the equation $\rd x_{t}=\mu_{t}\rd t+\sigma_{t}\rd W_{t}$, where $\{\mu_{t}\}$ 
and $\{\sigma_{t}\}$ are called the drift and the volatility of $\{x_{t}\}$. 
We can consider an Ito process $\{x_{t}\}$ taking values on a manifold 
${\mathfrak M}$, driven by
an $m$-dimensional Wiener process $\{W_{t}^{i}\}_{i=1,\ldots,m}$.
Let $\nabla_{a}$ be a torsion-free connection on ${\mathfrak M}$
such that for any vector field $\zeta^a$ its covariant derivative in
local coordinates is
\[
\delta_{\bf b}^b \delta_a^{\bf a}(\nabla_b\zeta^a) = \frac{\partial
\zeta^{\bf a}}{\partial x^{\bf b}} + \Gamma_{\bf bc}^{\bf a}\zeta^{\bf c},
\]
where $\delta_a^{\bf a}$ is the standard coordinate basis in a given
coordinate patch. Suppose we have an Ito process taking values in
${\mathfrak M}$. Let $x_t^{\bf a}$ denote the coordinates of the
process in a particular patch. Then writing $h^{\bf ab}
=\sigma_i^{\bf a}\sigma^{{\bf b}i}$ we define the drift process
$\mu^{\bf a}$ by
$\mu^{\bf a}\rd t= \rd x^{\bf a} +\half \Gamma_{\bf bc}^{\bf a}
h^{\bf bc} \rd t - \sigma_i^{\bf a}\rd W_t^i$. 
Alternatively, we can write the \textit{covariant Ito differential}
as $\rd x^a = \delta_{\bf a}^a ( \rd x^{\bf a}+\half \Gamma_{\bf
bc}^{\bf a} h^{\bf bc} \rd t)$, where $\delta^a_{\bf a}$ is the dual
coordinate basis. Then we have 
\[
\rd x^{a}= \mu^{a} \rd t+\sigma^{a}_{i}\rd W^{i}_{t} 
\]
for the covariant Ito differential on ${\mathfrak M}$ 
associated with the given connection.

\end{tcolorbox}

The important point is that while $\pi_{it}$ has no drift, the induced process on 
${\mathfrak M}$ has a drift that is given by the negative gradient of the variance 
of $X$. In other words, there is a tendency to flow towards the directions with 
zero uncertainty about the choice in question. 

What this means is that in 
assessing noisy information, people are drawn into states with lower 
uncertainties. In an extreme situation whereby a person is almost certain that 
one of the alternatives is the right choice, this state of zero uncertainty acts as 
a trap. That is, even if the person obtains noisy information that suggests that 
this is the wrong choice, a rational person who follows Bayesian updates will 
hardly change their view on the matter for a very long time (figure~3). 
I called this effect 
a `tenacious Bayesian' behaviour (Brody 2022). 

Today’s politics plays out against a backdrop of uncertainties that include 
wars in Ukraine and Gaza; the cost of living crisis; energy, 
food and water insecurity; migration; and so on. Above all, the impact of the 
climate crisis. The electorates are more than ever longing for certainties. 
This is why they are drawn into candidates who can offer a sense of certainty, 
irrespective of whether it is true or false. The rise of `populist' parties around 
the world, who offer simple solutions to complex societal challenges, when 
there are no simple solutions available, is a manifestation of this effect. 
Failing to realise the importance of 
this effect --- an effect that becomes clearly visible when applying stochastic 
differential geometry --- in political messaging, a candidate is unlikely to 
succeed in an election.

\section*{Modelling the impact of `fake news'}

\begin{wrapfigure}{r}{0.50\textwidth}
\includegraphics[width=0.95\linewidth]{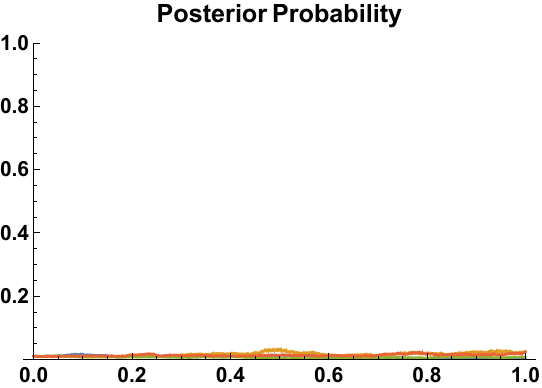}  
\caption{If $p_1=0.01$ and $p_2=0.99$, then even if the noisy information 
is provided by $\xi_t=\sigma x_1 t + \epsilon_t$, indicating that $X=x_1$ is 
the correct alternative so that $\pi_{1t}$ should approach $1$, a Bayesian 
thinker will maintain the position $\pi_{1t}\approx0$ for a long time. 
Four sample paths are shown here for $\sigma=1$.} 
\label{fig:3}
\end{wrapfigure}

Because we have a formalism to model how the flow of information affects 
people's behaviour, we can include a disinformation term to study its impact. To 
this end, note that a genuine noise (rumours, speculations, 
and so on) carries no unknown bias so that we can assume ${\mathbb E}
[\epsilon_t]=0$, because any known noise bias will be discarded. Hence 
if there is an unknown bias in the noise, this will skew the assessments of the 
receiver of the information. With this in mind, a simple model for disinformation 
can take the form  
\[
\xi_t = \sigma X t + (\epsilon_t + f_t) \, , 
\] 
where ${\mathbb E}[f_t]\neq0$ and where the existence of the term $f_t$ is 
unknown to the receiver of the message. 

With the help of the model we can conduct a quantitative statistical analysis of 
the impacts of disinformation on an election. For example, we can determine 
the averaged percentage change in the election outcome if disinformation of a 
given magnitude and frequency were disseminated. One result that might come 
as a surprise is that if the receiver is aware of the statistical distribution of the 
term $f_t$, then this is sufficient to almost entirely eradicate its impact (Brody 
\& Meier 2022). 
Fact checkers are helpful in determining whether any given information is 
true or false. But a more useful information is the detailed statistics of their 
findings, which is not commonly available. Mathematical analysis shows that 
such information is helpful to counter the impacts of disinformation in 
democratic processes, but currently, policy makers seem unaware of the fact that 
such mathematical analysis can be applied to develop countermeasures to tackle 
the issue. 

There is another situation in which an effective countermeasure can be 
implemented when right conditions are met. This concerns how information 
are combined. If there are two information sources, $\xi_t^1=\sigma_1 X t 
+ \epsilon_t^1$ and $\xi_t^2=\sigma_2 X t + \epsilon^2_t$, on the same issue 
$X$, then the combined effect is equivalent to the consumption of a single 
information source 
\[
\xi_t = \frac{\sigma_1-\rho\sigma_2}{\sigma(1-\rho^2)}\, \xi_t^1 + 
\frac{\sigma_2-\rho\sigma_1}{\sigma(1-\rho^2)} \, \xi_t^2 \, ,
\]
where $\rho$ is the correlation of the two Brownian noises $\epsilon^1_t$ and 
$\epsilon^2_t$, and $\sigma$ is a certain positive function of the parameters 
$\sigma_1$, $\sigma_2$, and $\rho$. We can think of $\xi^1_t$ and $\xi^2_t$ 
representing, for instance, information emerging from two competing political 
parties. Suppose that $\rho>0$ and that $\xi_t^2$, 
say, contains disinformation. In this case, by letting $\sigma_1>\sigma_2/\rho$, 
it is possible to reverse the impact of disinformation, that is, fake news will 
turn against its spreader (Brody \& Yuasa 2025). 
The lesson for the first political party therefore is to 
deliver clear and transparent messages to enhance $\sigma_1$, rather than 
to pay too much attention to counter the disinformation.

\section*{A pathological liar?}

What if you have a prime minister, or a president, who is a pathological liar? 
I shall not get into clinical definitions of a pathological liar here, but instead 
interpret it to simply mean 
a person who knows the fact but is determined not to reveal it for one reason or 
another. With this understanding, can mathematics uncover statistical signs of 
the behaviour of such a liar? The information-based approach is sufficiently 
versatile to give us an affirmative answer. 

\begin{wrapfigure}{r}{0.50\textwidth}
\includegraphics[width=0.95\linewidth]{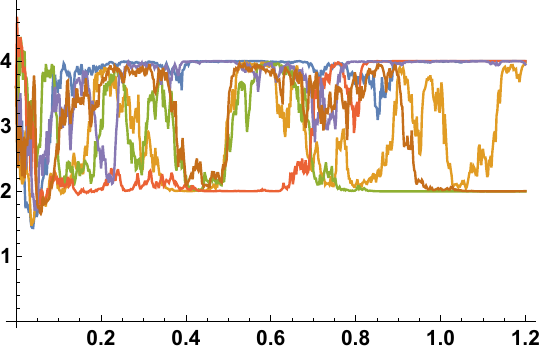}  
\includegraphics[width=0.95\linewidth]{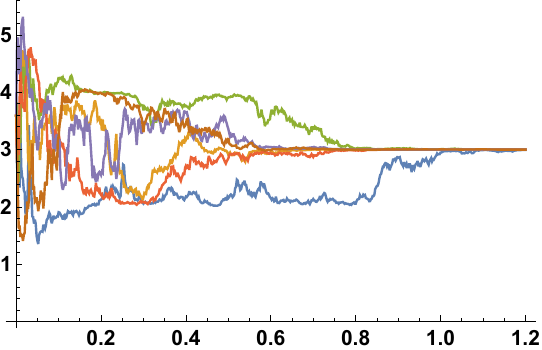}
\caption{There are six alternatives labelled by $X=1,2,3,4,5,6$. If $X=3$ but 
$p_3=0$, then the mean $X_t$ tends to hop between the closest alternatives 
$X=2$ and $X=4$, when there is a lot of ($\sigma=5$) 
reliable information coming out 
saying that $X=3$ (top panel).  
In contrast if $p_3=0.01$ but $X=3$, 
a genuine misunderstanding, then $X=2$ and 
$X=4$ remain 
plausible alternatives for a while, but the view will gradually converge to the true 
alternative $X=3$ (bottom panel). 
Six sample paths are shown in each case.} 
\label{fig:3}
\end{wrapfigure}

Recall that the posterior probabilities $\pi_{it}=\pi_i(t,\xi_t)$ are functions of two 
variables, time $t$ and information $\xi_t=\sigma X t + \epsilon_t$. One of the 
advantages of the 
information-based approach is that numerical simulations are very easy to 
implement because 
there is no need to solve stochastic differential equations. Specifically, we sample 
a value of $X$ and a Brownian path $\epsilon_t$, substitute them in $\xi_t$ and 
plug that into $\pi_i(t,\xi_t)$ to generate a simulation outcome, for example, of the 
expected mean $X_t=\sum_i x_i \pi_{it}$. The sampling of 
$X$ is done according to the prior probabilities $\{p_i\}$. 

Now suppose that $p_k=0$ for one of the alternatives. It is nonetheless possible 
to consider the sample $X=x_k$; an event that occurs with zero probability, and 
study the behaviour of the conditional mean $X_t$. When there is a lot of 
reliable information coming out saying that $X=x_k$ is the correct alternative, 
the result is quite revealing. 
The process $X_t$ rather abruptly converges to one of the `false' alternatives 
$x_l$, $l\neq k$, and stays there for a while. But noisy information being 
revealed 
suggests that $X=x_k$, so eventually $X_t$ is kicked off from the level $x_l$ and 
quickly converges into another false alternative $x_m$, and repeat this procedure 
by hopping from one false alternative to another, typically between the closest 
false alternatives $x_{k+1}$ and $x_{k-1}$ to the true alternative $x_k$ 
(figure~4). 

In other words, a pathological liar of the kind defined above would quite 
confidently pick one of the false alternatives and maintains the narrative, until 
enough information is revealed to make that position uncomfortable. At this 
point a different false alternative is quickly and confidently chosen, as if there is 
no inconsistency with the previous position, until when he is forced to change his 
position again. 

The behaviour of the process $\{X_t\}$, which is something reminiscent of a 
piecewise step function, when $p_k=0$ but $X=x_k$, is statistically 
distinguishable from the behaviour of $\{X_t\}$ when $p_k\ll1$ but is nonzero, 
no matter how small $p_k$ may be. The latter represents a situation in which 
a person 
is genuinely unaware that $X=x_k$ is the correct choice. So from the behaviour 
it is possible, at least with some statistical confidence, to determine whether 
someone is refusing to admit the truth (by blocking out the true alternative 
$X=x_k$) or is genuinely unaware of the truth, in an 
environment where noisy information about the truth is revealed.  

\section*{Transmission, feedback, and control}

Thus far our attention has been focused on how noisy information is processed 
and analysed, which we used to deduce the resulting action. Given the fact that 
this analysis generates an uncertainty-minimising flow, it is natural to ask whether 
an individual can transmit information back into the system in the form of a 
feedback, with the view to further reduce uncertainties. That is, we can think of 
the possibility of feedback control. To this end consider an individual (the system), 
the environment (community, society, or a section of the society), and the 
communication channel (Internet, Twitter / X, Instagram, TikTok, and so on). 

It is a well-known fact that channel capacity cannot be increased by feedback. 
Nevertheless, it turns out that noise-free feedback from the system to the 
environment can enhance the information gained from the environment. 
Consider the following, somewhat idealised setup of Davis (1978). In this setup, 
the information entering the system from the environment is obscured 
by noise, but the environment can detect the signal transmitted from the system 
without noise. The system (individual) is interested in the amount of information 
contained in the message coming from the environment about the value of the 
signal. This quantifier is known in communication theory as mutual information 
(Gelfand \& Yaglom 1957), which is determined by the likelihood ratio for 
detecting the signal. The question therefore is what feedback should the 
individual transmit back into the community so as to maximise the mutual 
information. In the case of a 
Gaussian signal, the optimal transmission turns out to be the increments 
$\rd W_t$ of the innovation process itself (Davis 1978). That is, only new 
information is transmitted from the system to the environment to 
maximally reduce the uncertainty of the signal. 

There is a profound implication of this result, albeit in a slightly idealised setup, 
in today's digital society. From individual's perspective, transmission of 
information is time and energy consuming. Hence in practice unless the 
magnitude $|\rd W_t|$ of the new information is above a threshold, it is unlikely 
that this is fed back into the community. Recall that the new information 
$\rd W_t$ is the discrepancy $\rd\xi_t - {\mathbb E}_t[\rd\xi_t]$ between the 
predicted and realised increments of the information. Now suppose that the 
information $\rd\xi_t$ over the time interval $[t,t+\rd t]$ is secretly 
contaminated with disinformation $\Delta f$. Then this will shift the realised 
increment $\rd\xi_t$ by $\Delta f$ but not the predicted increment 
${\mathbb E}_t[\rd\xi_t]$, because the receiver is unaware of the existence of 
disinformation. It follows that for sufficiently large $\Delta f$, the likelihood of 
the distorted innovation $|\rd W_t+\Delta f|$ exceeding the threshold value is 
very high. 

It is known empirically that the likelihood of a person disseminating disinformation 
is significantly higher than that of the person disseminating true information 
(Vosoughi \textit{et al}. 2018). The observation here demonstrates the intuitively 
acknowledged fact that the higher the novelty of information, the more likely it is 
that the information is transmitted. Importantly, though, this is the consequence of 
a highly rational behaviour, following the rule of optimal feedback control. The 
consequence therefore is to further enhance noise in the system, and this 
enhancement is amplified in a society that is becoming more reliant on the use of 
artificial intelligence systems.

\section*{Political polarisation}

The tenacious Bayesian behaviour shows that if a person is rational in the sense 
of processing information in accordance with the Bayes formula, then in the event 
of that person believing strongly in a false alternative, the scope of changing the 
mind of that person by supplying factual information is limited. It is then reasonable 
to ask whether an alternative approach might exist to deal with such circumstances 
whereby misguided people holding strong but undesirable political views. 

It is often pointed out that politics is becoming more and more polarised, but the 
same is applicable to society as a whole. To understand this phenomenon, it will 
be useful to cite Wiener's argument (Wiener 1950) that the effects resulting from the 
second law of thermodynamics on society cannot be overstated. The second law 
says that over time, order will be replaced by confusion and uncertainty -- a 
phenomenon that is inevitable in a society with rapidly-developing communication 
technology. 

A rational individual, on the other hand, seeks order and certainty. Indeed, one 
can show that the Bayesian updating follows from the requirement of maximal 
reduction in the expected uncertainty (Brody \& Trewavas 2023). 
That is, when we acquire noisy 
information $\xi$ about a quantity of interest $X$, there are in general uncountably 
many transformations from the prior view $p_i$ to the posterior view $\pi_i$ that 
are all consistent with the acquired information. But if we demand that the 
transformation should minimise, on average, the uncertainty of $X$, then the 
result agrees with the expression obtained by use of the Bayes formula for the 
conditional probability. Biologically this makes a lot of sense. If $X$ represents 
the unknown condition of the environment, for instance, 
then a protocol that maximally 
reduces uncertainties of $X$ is highly advantageous for survival because 
a biological system must adapt to the condition of its environment. 
Thus we have a conflict: The 
entropy that measures the degree of uncertainty 
of the society is rapidly increasing, but individually we try to reduce 
entropy. A corollary is that to keep the level of entropy low in such an 
environment, a person must limit the latitude of information source, resulting in 
the creation of echo chambers. The phenomenon of social polarisation therefore 
is the result in part of the competition between the second law and the individual 
desire to counter the second law. It should be evident that 
recent advances in artificial intelligence can 
only enhance this conflict. 

The situation is further exacerbated by the fact that unlike in mathematics or in 
physical 
science where the validity of a proposition can be tested, either by a mathematical 
proof or by repeatable experiments, in politics or in social 
science more generally, this is often not possible. The long-term consequence of 
an economic policy on society, for instance, cannot be determined with certainty. 
One might have an opinion on the matter, but it remains an opinion, no matter 
how informed that opinion might be. Hence the boundary of fact and fallacy is 
a fuzzy one in politics, making it harder for even a well-informed opinion to be 
heard by political opponents.

\section*{Incompatible propositions}

Given that the effect of the second law cannot be eradicated, the issue of 
political polarisation is a challenging one to address. Nevertheless, when 
there is an unambiguous fact available, for instance in the case of an event 
that had actually taken place, it is legitimate to ask whether a misguided 
individual strongly believing in a false alternative can be influenced. 

To explore this possibility, it is worth remarking that the Bayes formula, which 
tells us how a prior view on a topic is updated in accordance with the 
acquired information, 
is applicable only if the prior and the posterior probabilities are 
absolutely continuous with respect to each other, in the sense that they share 
a common null set. When the continuity condition 
is lost, or more generally when a sequence of information variables does not 
admit a joint probability, the Bayes formula is not applicable. However, even 
in those circumstances, an uncertainty-minimising transformation does exist, 
and is given by what is known as the L\"uders transformation. The idea can 
most easily be explained by relying on geometric intuitions and consider the 
square roots $\sqrt{p_i}$ and $\sqrt{\pi_i}$ of the prior and posterior probabilities. 

The vector space in which square-root probabilities are defined, when endowed 
with a Euclidean inner product, is known as a Hilbert space. This is the Hilbert 
space 
${\cal H}_{\cal S}$ that models our view of the `signal' $X$. There is likewise a 
Hilbert space ${\cal H}_{\cal N}$ that models our view of the `noise' $\epsilon$.
The tensor-product space ${\cal H}_{\cal S}\otimes{\cal H}_{\cal N}$ then models 
the space of all possible square-root joint probabilities of $X$ and $\epsilon$, 
where random variables on Hilbert space are modelled by matrices. 
When a noisy information is acquired, that is, when the value of $\xi=\sigma X 
+\epsilon$ is detected, this singles out a linear subspace ${\cal H}_{\xi}$ of 
${\cal H}_{\cal S}\otimes{\cal H}_{\cal N}$ corresponding to all possible 
combinations of the random variable pair $(X,\epsilon)$ such that the value of 
$\sigma X+\epsilon$ agrees with the detected value of $\xi$. The situation is 
best understood by considering the case in which $X$ is discrete and 
$\epsilon$ is continuous. Then given the value of $\xi=y$, we know either 
$X=x_1$ and $\epsilon=y-\sigma x_1$, or $X=x_2$ and $\epsilon=y-\sigma 
x_2$, or $X=x_3$ and $\epsilon=y-\sigma x_3$, and so on --- all these 
possibilities are consistent with the data $\xi=y$. The idea of the 
L\"uders rule is that the square-root of the \textit{a priori} joint probability is 
projected down onto ${\cal H}_{\xi}$ to form the square-root of the 
\textit{a posteriori} joint probability. When the result obtained is squared, and 
suitably normalised, we recover the Bayes formula. 
The geometric formulation thus uncovers the fact that not only the transformation 
from the prior to the posterior probabilities maximally reduces uncertainties but 
also it generates the minimum change in the probability that is consistent with 
the acquired information (Brody \& Trewavas 2023). This, again, is biologically 
advantageous because the smallest change in the view of the external world 
that is consistent with detected data leads to the smallest waist in energy 
consumption. 

Now suppose that having acquired the information $\xi=\sigma X +\epsilon$ one 
subsequently acquires another information, which is modelled by $\xi'=\sigma Y 
+\epsilon'$, where the two random variables $X$ and $Y$ do not admit a joint 
probability. In this case, the Bayes formula is not applicable, whereas the 
L\"uders projection rule is always applicable. When the two random variables 
$X$ and $Y$ do not admit a joint probability, we say that the two propositions 
modelled by these random variables are not compatible. Practically, what this 
means is that the matrix representations of $X$ and $Y$ on Hilbert space are 
not commutative. Thus we enter into the domain of \textit{noncommutative 
communication theory}. Such a seemingly abstract concept turns out to be 
highly relevant in political messaging. 

In cognitive psychology, for long it was taken for granted, quite understandably, 
that all propositions are compatible. One consequence of this assumption is that when people's behaviour deviates away from what is predicted by the rules of 
probability, for example a violation of the Bayesian updating, such a behaviour 
can be classified as irrational. A classic example is the notorious conjunction 
fallacy (Tversky \& Kahneman 1983) , whereby ``$A$ and $B$'' is thought more 
likely than``$A$". According to standard logic and probability theory this is never 
true, but in experiment after experiment it can be human-true. But standard logic 
and probability theory is applicable only so far as when $A$ and $B$ admit a 
joint probability law. 

It is legitimate therefore to ask: What if $A$ and $B$ were not compatible? This 
is in effect the question addressed by Jerome Busemeyer and others in 
cognitive science (see Busemeyer \& Bruza 2024, and references cited therein). 
The answer they have found was remarkable. Much of the seemingly 
irrational human behaviours reported in the cognitive psychology literature, for 
which there are no straightforward explanations using only compatible 
propositions, could be explained consistently using models accommodating 
incompatible propositions. Statistical analysis of random events involving 
incompatible propositions, however, is what quantum theory is about. Indeed, 
there are two fundamental 
distinctions between Kolmogorovean classical probability theory 
and what is 
known as quantum probability theory. The first is that the square-root probabilities 
$\psi_i$, called probability amplitudes in quantum mechanics, can take complex 
values so that the probabilities are recovered using modulus squares: 
$p_i=|\psi_i|^2$. (Imaginary number plays two roles in quantum mechanics, 
namely, to rotate plane waves by the right angle, and to set the direction of 
time. It is known that quantum theory cannot be formulated without the use of 
complex numbers, whereas it is not known whether complex numbers are 
required for a consistent characterisation of cognitive behaviours.) The second 
is that random variables, which are modelled on Hilbert space as matrices, are 
in general not compatible. With the realisation that quantum probability models 
can more accurately and succinctly model cognitive behaviours, a new field 
emerged within cognitive science called `quantum cognition', which is now an 
active field of research. 

It may be added parenthetically that quantum cognition is not concerned with 
quantum effects that may or may not be taking place at the neurophysiological 
level. Rather, it concerns the application of quantum probability assignment rules 
to model cognitive human behaviour. If a pattern of human behaviour appears to 
violate 
rules of classical probability but respects quantum rules, then one can no longer 
classify this as being irrational, because when there are incompatible propositions, 
classical probability is simply not applicable to say what the rules are. The 
situation is entirely analogous to the discussion concerning the famous Bell inequality in 
quantum mechanics (Isham 1995). The bound on a certain correlation in the Bell 
inequality is obtained by assuming the existence of a joint probability distribution, 
so when there is no such distribution, classical probability is simply not applicable. 
It is then illogical to argue that a violation of the Bell inequality is a violation of 
classical laws of physics, because classical physics provides no bound. Where 
it violates classical physics is the existence of incompatible propositions 
itself, and not the inequality. In a similar 
vein, if a seemingly irrational behaviour is consistent with the rules of quantum 
probability, then this should be viewed as entirely rational, because quantum rules 
provide a high level of efficiency that should be preferred biologically 
(Brody \& Trewavas 2023). 

The relevance of what I call here noncommutative communication theory in politics 
has long been appreciated by seasoned politicians and political advisors, purely 
on the empirical or intuitive grounds, as an art of messaging. Perhaps a simple 
anecdotal example suffices to highlight this. Take the 2016 `Brexit' referendum 
in the United Kingdom on membership of the European Union. Some `leave' 
campaigners put to the public the question, before the referendum, whether 
they want more funding for the National Health 
Service, with the implication, true or false, 
that leaving the European Union leads to more 
funding for the health service. If the membership question and the health service 
funding question were compatible, then classical probability theory shows that 
asking the funding question will have no implication in the referendum outcome. 
Those leave campaigners believed otherwise, and the result was `out'. 

There are innumerable such examples whereby political campaigners 
instinctively know how to shift the frame of reference of the public opinion, 
but the science underpinning this effect is only emerging in the form of 
quantum cognition and noncommutative communication theory, over the 
past two decades. Indeed, 
even in the hypothetical scenario whereby a person believes 100\% in an 
alternative, there is a way of changing the view of that person -- an impossibility 
in classical probability (Basieva \textit{et al}. 2017). 
The trick is not to convince the person otherwise, 
which is futile because a counterargument to that alternative is compatible 
with the argument itself. Counterintuitive this may be, a counterargument 
will have no effect on the view of that person. 
Instead, by asking a single incompatible question, and importantly 
a single question is 
sufficient here, it is possible to change the belief of the person on the original 
issue to a number strictly less than 100\%.

\section*{Discussion}

As a discipline, mathematical politics in the sense of an art of messaging is very 
much at the stage of infancy. While 
much insights have been gained with perhaps the simplest nontrivial models, there 
are scopes for further developments. Indeed, mathematical theory underpinning 
signal detection, sometimes called `stochastic filtering theory', has developed 
considerably over the past 60 years. It seems inevitable that some of the more 
powerful machineries can be applied to gain a wide range of new insights in 
political messaging and beyond. 

There are new mathematical challenges that 
emerge as well. For example, in traditional communication theory, there are no 
ambiguities as for the interpretation of the signal (the random variable $X$ in the 
current example). In politics, however, there is a huge ambiguity. Consider a 
newspaper article on an issue. The view of a person on this issue may be modelled 
by $X$ (taking the values $\{x_i\}$ with the probabilities $\{p_i\}$) so that the 
information contained in the article can be modelled, say, by $\xi=\sigma X+
\epsilon$; but only for that person. Another reader of the \textit{same} 
article will have a different view on the issue, represented by $Y$ 
(also taking the values $\{x_i\}$ but with the probabilities $\{q_i\}$), so the 
information model for that person takes the form, say, $\xi'=\sigma Y+\epsilon$, 
even though we are speaking about an identical news article. Thus, 
mutual information of a message, that is, the amount of information contained 
in the message about the value of the signal, is receiver dependent. In a time 
series version of the problem, statistical analysis requires the specification of 
information in the form of conditioning, but when there are ambiguities regarding 
which, or whose information to consider, it is not clear how to conduct such 
analysis, and this is a challenge already in pure mathematics. 

Beyond politics, the programme introduced here can have profound implications 
in other areas. One is in cognitive psychology. For instance, the 
tenacious Bayesian phenomenon discussed above can be used to show 
that some of the common conclusions in the psychology literature on the 
seemingly irrational behaviour of people are not always correct (Brody 2022). 
Mathematical analysis, when applied properly, offers so much potential to model 
and explain cognitive behaviours, both qualitatively and quantitatively. When this 
is combined with developments in quantum cognition and noncommutative 
communication theory, the scope of a better understanding of human 
behaviour, in the cognitive context, broadens further. At the same time, 
one should not underestimate the risks involved, for, a better understanding 
of our cognitive response to signals can potentially be used to control people, 
both in positive and negative ways. 

Another application outside politics is in advertisement and marketing. Because 
the mathematical theory of politics is concerned with the understanding of the 
impacts of political messaging, the findings can be extended, 
\textit{mutatis mutandis}, to marketing. This is because the questions 
being addressed are very similar in the two fields --- how to maximise the 
impact of a message, either in the form of vote share or of the sales figure. In 
both areas, more traditional regression analysis is commonly employed, but 
the mathematical theory of information and communication can offer 
considerably more towards generating future statistics. 

It is of interest also to explore behaviours of artificial intelligence 
from the viewpoint of the present discussion. The mathematics of messaging 
is concerned with understanding how a signal is processed by an intelligent 
system. Presumably today's artificial intelligence would process signals differently 
from what the theory predicts, 
but if so, how different are they? It is hoped that a better understanding of the 
mathematics of messaging will lead to an alternative and better design of an 
artificial intelligence architecture 
that is not only transparent but also is significantly more 
energy efficient --- an endeavour that is not too dissimilar to the attempts based 
on the framework of active inference (Pezzulo \textit{et al}. 2024). 

The idea of modelling information surrounding us as a probabilistic 
time series may at first seem abstract and intangible. As Edgar Allan 
Poe wrote, 
``Such sentiments are seldom thoroughly stifled unless by reference to the 
doctrine of chance, or, as it is technically termed, the Calculus of Probabilities. 
Now this Calculus is, in its essence, purely mathematical; and thus we have 
the anomaly of the most rigidly exact in science applied to the shadows and 
spirituality of the most intangible in speculation.'' The information-based 
approach to mathematical politics shows that the matter is not quite so 
intangible as one might have thought at first.



\end{document}